\documentclass[twocolumn]{aastex63}

\received{January 20, 2021}
\accepted{February 17, 2021}
\submitjournal{ApJ}

\shorttitle{Cepheids with giant companions. I.}
\shortauthors{Pilecki et al.}
\graphicspath{{./}{figs/}}
\begin{document}

\title{Cepheids with giant companions. I. Revealing a numerous population of double-lined binary Cepheids
\footnote{Based on observations collected at the European Southern Observatory, Chile}
\footnote{This paper includes data gathered with the 6.5m Magellan Clay Telescope at Las Campanas Observatory, Chile.}
}

\correspondingauthor{Bogumił Pilecki}
\email{pilecki@camk.edu.pl}

\author[0000-0003-3861-8124]{Bogumi{\l} Pilecki}
\affiliation{Centrum Astronomiczne im. Miko{\l}aja Kopernika, PAN, Bartycka 18, 00-716 Warsaw, Poland}

\author{Grzegorz Pietrzy{\'n}ski}
\affiliation{Centrum Astronomiczne im. Miko{\l}aja Kopernika, PAN, Bartycka 18, 00-716 Warsaw, Poland}

\author{Richard I. Anderson}
\affiliation{Institute of Physics, Laboratory of Astrophysics, EPFL, Observatoire de Sauverny, 1290 Versoix, Switzerland}
\affiliation{European Southern Observatory, Karl-Schwarzschild-Str. 2, 85748 Garching b. München, Germany}

\author{Wolfgang Gieren}
\affiliation{Universidad de Concepci{\'o}n, Departamento de Astronom{\'i}a, Casilla 160-C, Concepci{\'o}n, Chile}

\author{M{\'o}nica Taormina}
\affiliation{Centrum Astronomiczne im. Miko{\l}aja Kopernika, PAN, Bartycka 18, 00-716 Warsaw, Poland}

\author{Weronika Narloch}
\affiliation{Universidad de Concepci{\'o}n, Departamento de Astronom{\'i}a, Casilla 160-C, Concepci{\'o}n, Chile}

\author{Nancy R. Evans}
\affiliation{Smithsonian Astrophysical Observatory, MS 4, 60 Garden St., Cambridge, MA 02138}

\author{Jesper Storm}
\affiliation{Leibniz-Institut für Astrophysik Potsdam, An der Sternwarte 16, D-14482, Potsdam, Germany}

\begin{abstract}

Masses of classical Cepheids of 3 to 11 M$\odot$ are predicted by theory but those measured, clump between 3.6 and 5 M$\odot$. As a result, their mass-luminosity relation is poorly constrained, impeding our understanding of basic stellar physics and the Leavitt Law.
All Cepheid masses come from the analysis of 11 binary systems, including only 5 double-lined and well-suited for accurate dynamical mass determination. We present a project to analyze a new, numerous group of Cepheids in double-lined binary (SB2) systems to provide mass determinations in a wide mass interval and study their evolution.
We analyze a sample of 41 candidate binary LMC Cepheids spread along the P-L relation, that are likely accompanied by luminous red giants,
and present indirect and direct indicators of their binarity. In a spectroscopic study of a subsample of 18 brightest candidates, for 16 we detected lines of two components in the spectra, already quadrupling the number of Cepheids in SB2 systems. Observations of the whole sample may thus lead to quadrupling all the Cepheid mass estimates available now. For the majority of our candidates, erratic intrinsic period changes dominate over the light travel-time effect due to binarity. However, the latter may explain the periodic phase modulation for 4 Cepheids.
Our project paves the way for future accurate dynamical mass determinations of Cepheids in the LMC, Milky Way, and other galaxies, which will potentially increase the number of known Cepheid masses even 10-fold, hugely improving our knowledge about these important stars.

\end{abstract}



\section{Introduction} \label{sec:intro}

Classical Cepheids (hereafter also {\em Cepheids}) are perhaps the most important objects in astrophysics, crucial for various fields of astronomy like stellar oscillations and evolution of intermediate and massive stars, and with enormous influence on modern cosmology.  Since the discovery of the relationship between their pulsation period and luminosity over a century ago (the Leavitt Law, \citealt{1912HarCi.173....1L}), the Cepheids are extensively used to measure distances both inside and outside of our Galaxy.
The recent local Hubble constant determination accurate to 1.9\% \citep{Riess_2019_Hubble_1.9} that shows a significant discrepancy with the value inferred from the Planck data \citep{Planck_2018_Hubble} highly depends on the aforementioned relation.

\begin{figure}
    \centering
    \includegraphics[width=1.01\linewidth]{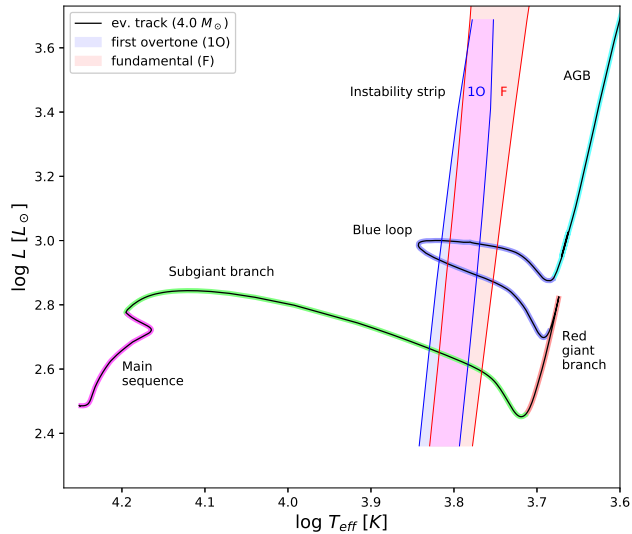}
    \caption{Hertzsprung-Russell diagram with the evolutionary track for a star of 4 $M_\odot$, which is typical for classical Cepheid variables. The instability strip for fundamental and first-overtone Cepheids is overplotted. Most Cepheids are found on the blue loop, but they also appear during the rapid stage of evolution on the subgiant branch.}
    \label{fig:HRdiag}
\end{figure}

Classical Cepheids are evolved intermediate and high mass, radially pulsating giants and supergiants.  Their well defined position on the helium-burning loop (called the {\em blue loop}; see Fig.~\ref{fig:HRdiag}) makes them also a sensitive probe of various properties important for evolutionary studies. The mass, metallicity, overshooting, mass loss and rotation affect significantly the shape and extent of the blue loop, determining the way and number the instability strip is penetrated. The Cepheids play also an important role in testing and development of pulsation theory \citep{Buchler_2009_AIPC}.

In the last decade the expected fraction of classical Cepheids in multiple systems has grown from about 30 to 80\% \citep{Evans_2015AJ_cep_binarity_30,Kervella_2019_Multiplicity}, indicating that we cannot ignore the multiplicity in our studies of these variable stars. The light of a companion directly affects their observed brightness, but the tidal deformation, mass transfer and the merger origin may have impact on their intrinsic brightness as well. The binary interactions, changing the stellar rotation, chemical composition, and mass may affect the evolution of Cepheids, and influence their pulsation characteristics. 

Apart from these complications, the multiplicity brings also some opportunities. Since the first measurement \citep{cep227nature2010} accurate masses of 6 classical Cepheids in double-lined eclipsing binary systems in the Large Magellanic Cloud (LMC) have been obtained \citep{cep227mnras2013,allcep_pilecki_2018}, helping in understanding of the Cepheid mass discrepancy problem\footnote{More information on the Cepheid mass discrepancy problem in \citet{Gieren_1989AA_cep_masses,Bono_2001APJ_cep_MR_mass_discrep}.} (between masses from evolution and pulsation theory; \citealt{Cassisi_Salaris_2011_ApJ}) and serving to test the pulsation theory models \citep{Marconi_2013_cep227}.
As only Cepheids accompanied by other giants (in a similar stage of evolution) were observed and analyzed, the measured mass ratios were mostly close to unity, but surprisingly for one system significantly different masses ($M_2/M_1$$\sim$0.7) were found. This led to a conclusion, that the system (OGLE-LMC-CEP-1812; \citealt{cep1812apj2011}) was a triple before, and the Cepheid is a product of a merger \citep{Neilson2015_cep1812_merger}. It is estimated that even 30\% of Cepheids may be such a product \citep{Sana_2012_BinInteractions}. Moreover, one of the studied eclipsing binary systems resulted to be composed of two Cepheids (OGLE-LMC-CEP-1718; \citealt{cep1718apj2014}).

Since the discovery of binarity of Polaris in 1929 \citep{Moore_1929_Bin_Polaris} many Cepheids in binary systems in the Milky Way were also identified \citep{Szabados_2003_BinCepCat}, but not a single one was found in an eclipsing or double-lined spectroscopic binary. The result is that in the Milky Way we know dynamical masses of only 5 classical Cepheids \citep{Evans_2018_V350SGr_Mass}, but they are mostly rough estimates from single-lined spectroscopic binaries, accurate typically to 10-20\%. For a few of them a great effort is taken to observe them from space in far-UV, where companion's lines are detectable and velocities can be measured \citep{Gallenne_2018_V1334Cyg_Cep}.
Please note, that in this paper we will keep calling a spectroscopic double-lined binary (hereafter also SB2) only objects for which lines of both components are easily detectable at {\em visual} wavelengths from ground observations.

Although theoretical studies of Cepheids are quite advanced \citep{Bono_1999_Cep_TheoryII,Bono_2005_ApJ_CEP_puls_model_IV,
Caputo_2005_ApJ_puls_evo_masses_CCep,Valle_2009_AA_theo_pred_CC_puls,Neilson_2012_AA_LD,Vasilyev_2017_AA_2dcep_model}, our observational data are still very limited.
Theory predicts masses of Cepheids in a range of 3-11 M$_\odot$ \citep{Cox_1980_ARAA_CEP_masses,Bono_1999_ApJS_cep_masses_range_high,Bono_2001_ApJ_cep_masses_range_low, Anderson_2016_Rotation} but their measured masses clump between 3.6 and 5 M$_\odot$, with only one higher but uncertain value of 6 M$_\odot$ \citep{allcep_pilecki_2018,Gallene_2019AA_Galactic_Cepheids,Evans_2018_V350SGr_Mass}. This makes the mass-luminosity relation very poorly constrained \citep{Anderson_2016_Rotation}, while it is crucial for the theoretical understanding of the period-luminosity (P-L) relation and the basic stellar physics regarding, e.g. the convection, mass-loss, and rotation. Moreover, the blue loops predicted by the evolution theory are too short to explain the existence of low-mass short-period Cepheids.

Masses in a wider range would be of utmost importance to solve these issues, but we can measure them (using Kepler's laws) practically only for Cepheids in spectroscopic double-lined binaries, for which lines of companions are easily detectable. Unfortunately, such systems are very rare. Most binary Cepheids (and all found in the Milky Way) have an early-type broad-line main sequence companion that is typically 2-5 mag fainter in the V-band \citep{BohmVitense_1985_ApJ_cep_blue_companions}, making it extremely hard to determine its velocity and thus the mass of the Cepheid.

One good source for more binary Cepheids would be to look for eclipses, but this solution has two serious limitations. First, only about a half of eclipsing systems are double-lined, and second, the best places to look for them are already examined and we do not expect to find many more such binaries soon. This is why we have to look for another, perspective source that was not yet considered and remains unexplored so far.

In this paper we present such a source describing its potential and we show the first results. In Section~\ref{sec:object_selection} we formulate the hypothesis regarding the Cepheid binary candidates and present the selection criteria. In Section~\ref{sec:data} we present the results from the analysis of photometric data. In Section~\ref{sec:spectroscopy} we show the results from the spectroscopic observations that prove the hypothesis presented before. In Section~\ref{sec:conclusions} we draw conclusions from the presented results and describe the next steps for this project.

\section{The hypothesis}
\label{sec:object_selection}

As described above, there is a great need for an independent source of Cepheids in binary systems that are well-suited for mass determination. The most valuable would be those in double-lined spectroscopic binaries, for which lines of both components are present in the spectra. 
To meet these conditions, one has to find Cepheids accompanied by stars of similar luminosity, and preferentially of late spectral types, i.e. at a subgiant or later stage of evolution. To identify them, we can consider at least three observable features caused by such companions:

\begin{itemize}
\item the total observed brightness of a Cepheid should increase significantly
\item its photometric pulsation amplitude (expressed in magnitudes) should decrease
\item its color should be either similar or redder (we expect companions mostly on the red giant branch or the blue loop)
\end{itemize}

\begin{figure}
    \centering
    \includegraphics[width=1.01\linewidth]{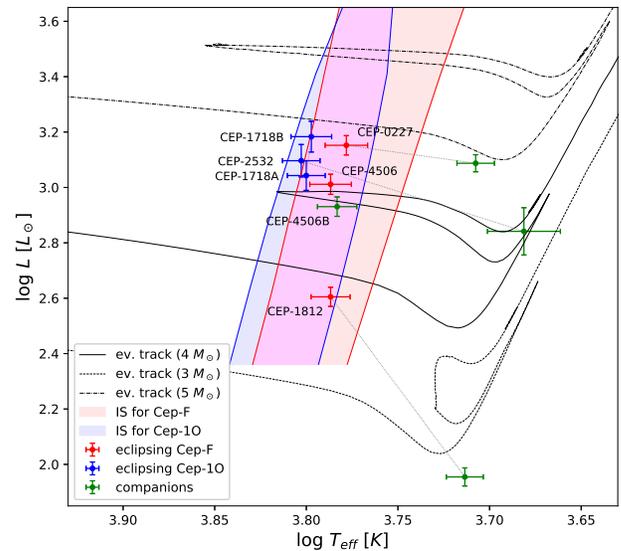}
    \caption{Positions of the components of known eclipsing binary systems with Cepheids on the HR diagram. BaSTI evolutionary tracks for 3, 4, 5 $M_\odot$, shifted by +0.15 in $\log L$ to match the measured Cepheid luminosities, are shown. It is evident that evolved companions can be mostly either similar or redder than Cepheids and of comparable brightness.}
    \label{fig:HRecl}
\end{figure}

We can test this assumptions looking at 5 similar SB2 eclipsing binary systems that were studied before \citep{allcep_pilecki_2018}. In Fig.~\ref{fig:HRecl} we can see their position on the Hertzsprung-Russell diagram with overplotted BaSTI evolutionary tracks \citep{Pietrinferni_2004ApJ_BaSTI} using canonical models for the LMC metallicity, and with the mass loss efficiency $\eta = 0.4$. The tracks were shifted by 0.15 in $\log L$ towards higher luminosities to match the Cepheids with corresponding masses\footnote{Non-canonical models give higher luminosities but predict blue loops too short to reach the instability strip.}.  The companions to the Cepheids are either redder or similar in color. In general they have also a similar brightness, except for the aforementioned OGLE-LMC-CEP-1812 system, which is probably of the merger origin. Its components have significantly different masses, yet they are found at a similar evolutionary stage. But even in this case the companion's brightness is about 50\% of that of the Cepheid.

\begin{figure*}
    \begin{center}
        \includegraphics[width=0.9\linewidth]{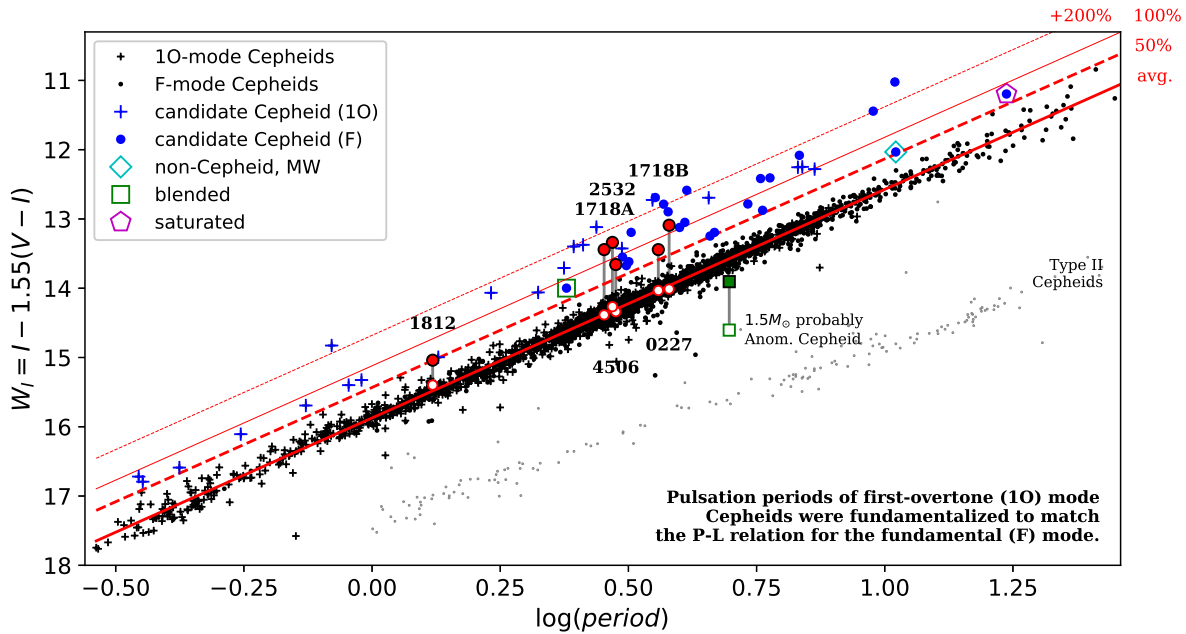} 
    \end{center}
\caption{P-L relation for reddening free Wesenheit index. Filled red circles show known eclipsing binaries with Cepheids (OGLE catalog numbers), while empty circles mark their brightness without the companion light.
Our candidates for binary Cepheids are $\ge$50\% brighter than an average Cepheid for its period. Note that their distribution parallel to the P-L relation advocates strongly against blending by a random star. Outlying points below the P-L relation may be binary type-II Cepheids and Anomalous Cepheids.}
\label{fig:perlum}
\end{figure*}

\begin{figure}
    \begin{center}
        \includegraphics[width=1.01\linewidth]{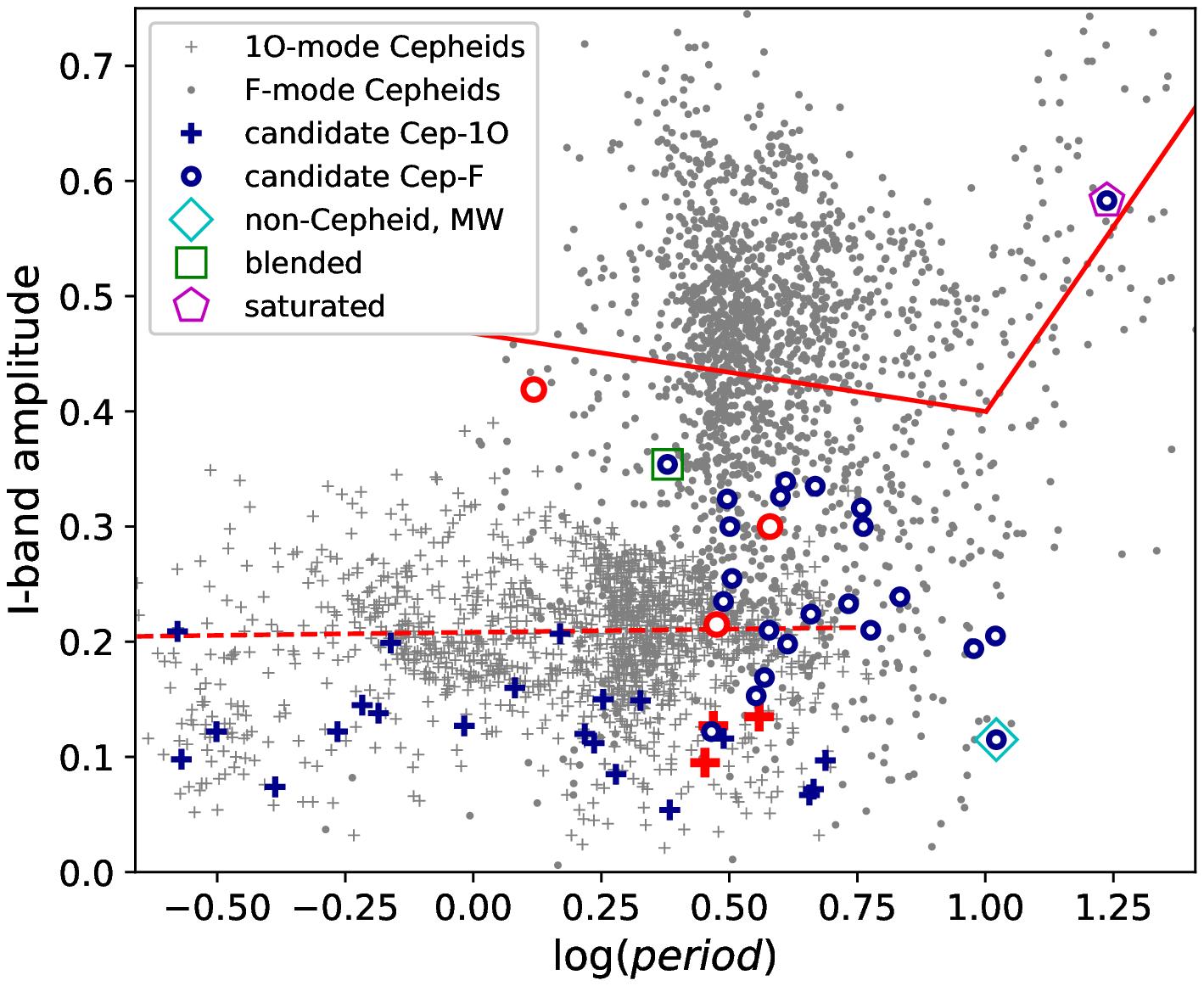} 
        \includegraphics[width=1.01\linewidth]{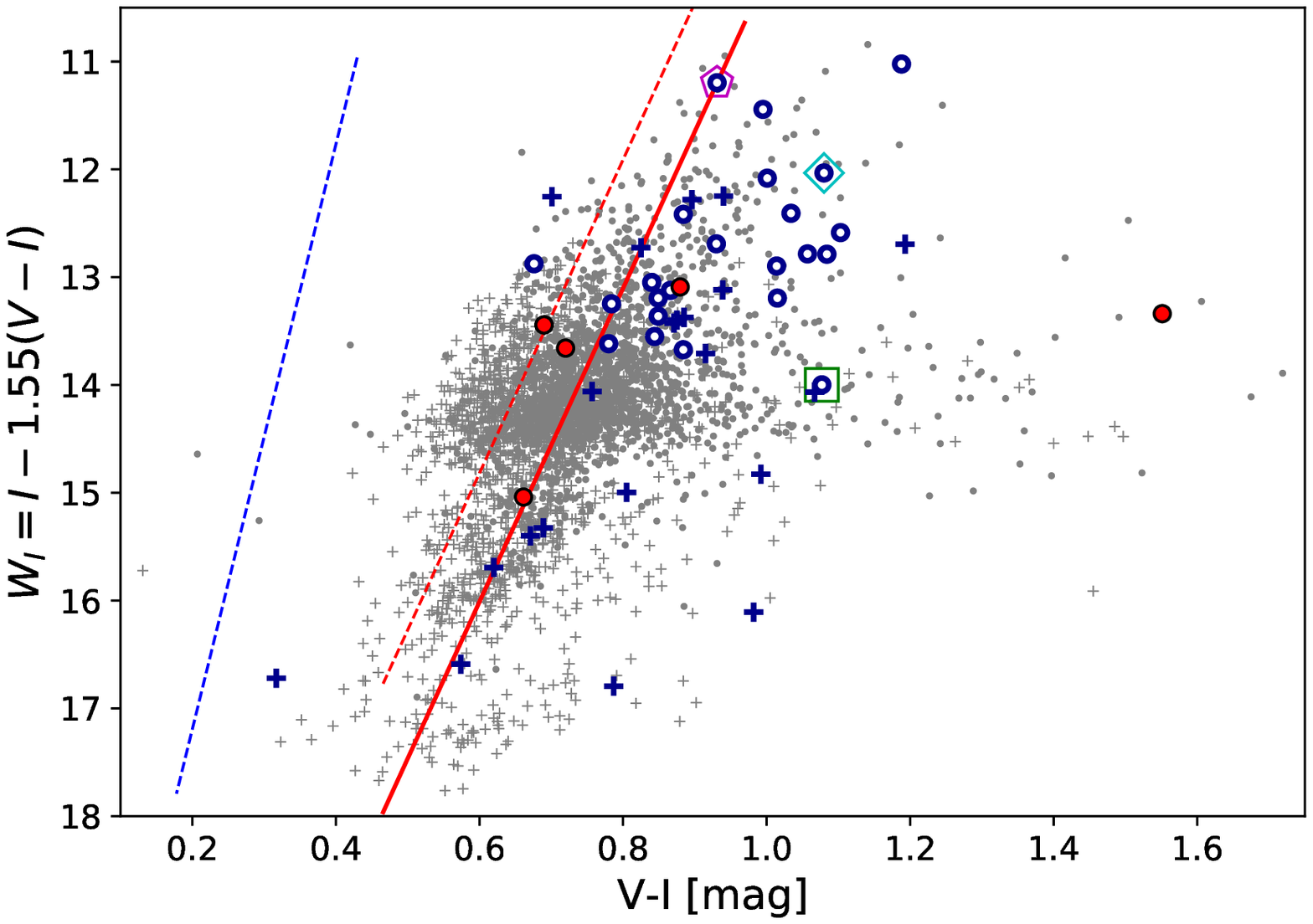}
    \end{center}
\caption{left: Period-amplitude diagram for classical Cepheids. Red lines mark average amplitude for each mode as a function of period. Virtually all our candidates are at the very low-amplitude end for their mode. right: Color-magnitude diagram. Red solid line marks a typical Cepheid color. As expected our selected objects are on average redder, especially if we take into account the effect a same-color, same-luminosity companion would have on a brightness of a Cepheid (red dashed line). Blue dashed line mark the effect of a blue companion of a similar brightness. Cepheids in known eclipsing binaries are marked with red circles and crosses (F- and 1O-mode, respectively) in the left panel, and with red circles in the right panel.}
\label{fig:peramp}
\end{figure}

Looking at the period-luminosity diagram for the LMC Cepheids (Fig.~\ref{fig:perlum}) one can see that all of the confirmed eclipsing giant-giant SB2 systems with Cepheids (red circles) lie significantly above the corresponding P-L relation, being at least 50\% (0.44 mag) brighter than a typical Cepheid for its period. After the subtraction of the companion's light, these Cepheids move close to the average P-L relation (empty circles).
In the left panel of Fig.~\ref{fig:peramp} (red circles and crosses) one can also see that their amplitude is about half of the typical one (red lines). The light dilution effect on the amplitude is smaller only for OGLE-LMC-CEP-1812 that has a significantly fainter companion. Note that there is a break in the values of Cepheids amplitudes at a period of about 10 days \citep{Klagyivik_2009_CEP_P-Amp}.

From the number of these eclipsing binary Cepheids we can also try to estimate how many similar binary Cepheids exist, but have not been detected in the photometric studies due to the lack of eclipses. The lowest inclination among the known eclipsing Cepheids is $i = 83^\circ$ (for OGLE-LMC-CEP-1718; \citealt{allcep_pilecki_2018}) at which the system shows a grazing eclipse. Assuming random inclinations, we can expect of the order of 50 SB2 systems composed of a Cepheid and a giant companion in a range of periods similar to the one for eclipsing Cepheids (i.e. 1-4 years). 

We investigated the P-L diagram for all the Cepheids from the OGLE catalog \citep{Soszynski_2008_clcep_cat} and we identified 44 more Cepheids that lie 0.44 mag (4.7$\sigma$) above the P-L relation, for which eclipses were not detected (Fig.~\ref{fig:perlum}). Virtually all Cepheids selected this way have low amplitudes (left panel of Fig.~\ref{fig:peramp}) and colors redder than typical, on average by 0.12 mag in V-I (right panel). As stated above, these three features together clearly indicate that such overbright Cepheids have luminous late-type giant companions. This, in turn, makes them perfect candidates for SB2 systems composed of giants, for which lines of all components can be easily detected and their radial velocities (RV) measured.
In Fig.~\ref{fig:perlum} periods of the first-overtone mode Cepheids ($P_{1O}$) were fundamentalized to match the periods of fundamental mode Cepheids ($P_{F}$) using the following formula:

\begin{equation}
P_F =  P_{1O} * (1.418 + 0.115 \log P_{1O}).
\end{equation}

\noindent This formula was derived by minimizing the scatter of the P-L relation for the combined set of all fundamental and first-overtone Cepheids, using the period $P_F$ for both and excluding the outliers.

The distribution of our candidate Cepheids parallel to the P-L relation confirms the hypothesis of a similar evolutionary stage of the companions and advocates strongly against blending which would statistically affect faint Cepheids more than the luminous ones. Also, blending would be rather random in colors, shifting stars toward redder (evolved companions) or bluer (main-sequence companions) colors, while practically all are red-shifted.  Only for 1 of 41 objects there is a possibility, that the shift in color is caused by a blue companion.

From that sample we excluded three objects, one that is saturated (OGLE-LMC-CEP-0535) and one that is marked as blended (OGLE-LMC-CEP-1583) in the OGLE catalog, as well as one object marked as possibly spotted (rotating) star (OGLE-LMC-CEP-0016). For the latter the light curve changes are more complex and we could not obtain a good fit, even taking into account the period change (see the next section).
Moreover, the distance inferred from the Gaia early Data Release 3 (eDR3; \citealt{Gaia_EDR3_2020_main}) parallax is about 2650 +/- 130 pc, pointing to its Galactic origin. This was also confirmed by its radial velocity of about 75 km/s, while values around 260 km/s are expected for LMC objects.
Excluding these three stars we were left with 41 promising candidates for binary Cepheids (see also Fig.~\ref{fig:perlum}).

The Cepheids in this sample are spread all over the P-L relation from 0.26 to 10.5 days. Among them 20 pulsate in the fundamental (F) mode and 21 in the first-overtone (1O) mode. Three of the latter are actually double-mode Cepheids pulsating in the second-overtone (2O) as well. It is also interesting that in the group selected this way there are two double Cepheids, i.e. objects where presence of two Cepheids is identified at the exactly same coordinates.
More details about these interesting and very valuable objects will be presented in the next paper of the series (Paper II, Pilecki et al., in prep.).

\section{Data analysis}
\label{sec:data}

\subsection{Periodicity}
\label{sub:periodicity}

We looked for periodicity of all selected objects using all the available data sets from the OGLE project \citep{Soszynski_2008_clcep_cat,Soszynski_2017AcA_OCVS_MC_Cep,ogle2015udalski}, collected with both I and V-band filter. We analyzed all the data sets at the same time, excluding only those that did not have enough data for a reliable fit with a Fourier series. For the same-band light curves from the OGLE-3 and OGLE-4 the amplitudes were fitted separately (as they may differ in general) but we kept the same phase coefficients. This way we limited the number of free parameters and we made sure that there is no phase shift between the datasets. This shift is an effect of a non-linear period variability, which is not averaged-out during the timespan of one dataset. When the same-band constraints on phase coefficients are not taken into account a sudden break may appear in the transition from OGLE-3 to OGLE-4 data.
The results of this analysis are presented in Table~\ref{tab:puls_ephem}.
For double-mode Cepheids the variabilities related to both modes were first disentangled and data only for the stronger first-overtone mode are shown in the table. The analysis and its results for two double Cepheids from the sample will be presented separately in Paper II.
In the photometric data of OGLE-LMC-CEP-3037 one eclipse is present, which suggest that this Cepheid may belong to an eclipsing binary system. The eclipse was removed before the periodicity analysis. This Cepheid is flagged pECL in the table.

\begin{deluxetable*}{cccccccc}
%\tablenum{1}
\tablecaption{Ephemeris time data for binary Cepheid candidates with giant companions \label{tab:puls_ephem}}
%\tablewidth{0pt}
\tablehead{
\colhead{} & \colhead{} & \colhead{} & \colhead{constant $P$} & \multicolumn2c{linear $dP/dt$} & \colhead{} & \colhead{}\\
\colhead{OGLE ID} & \colhead{mode} & \colhead{$T_0$ [days]} & \colhead{period [days]} & \colhead{period[days]} & \colhead{$dP/dt$} & \colhead{I [mag]} & \colhead{flags}
}
%\decimals
\startdata
LMC-CEP-0973 & F & 6083.2990 & 10.46505(8) & 10.46506(9) & -6e-08 [0.2$\sigma$] & 12.86 & - \\
LMC-CEP-0751 & F & 6001.1545 & 9.49519(8) & 9.49531(10) & -3.9e-07 [1.7$\sigma$] & 12.99 & QPOC \\
LMC-CEP-1957 & F & 3178.7704 & 6.814157(5) & 6.814157(5) & -1.6e-12 [0.0$\sigma$] & 13.63 & - \\
LMC-CEP-2096 & F & 5296.5907 & 5.971977(6) & 5.971971(6) & -1.7e-08 [2.2$\sigma$] & 14.01 & - \\
LMC-CEP-0837 & F & 4688.8020 & 5.7773217(26) & 5.777312(3) & -1.4e-08 [4.3$\sigma$] & 13.92 & - \\
LMC-CEP-0584 & F & 3361.7969 & 5.725339(19) & 5.725344(19) & -7.5e-08 [1.3$\sigma$] & 13.79 & - \\
LMC-CEP-2999 & F & 5848.9308 & 5.406430(7) & 5.406460(9) & 4.3e-08 [4.9$\sigma$] & 14.42 & - \\
LMC-CEP-1472 & 1O & 4747.0853 & 4.871930(7) & 4.871933(9) & 3.5e-09 [0.5$\sigma$] & 13.67 & - \\
LMC-CEP-1077 & F & 5228.0243 & 4.654260(3) & 4.654257(4) & -7e-09 [1.6$\sigma$] & 14.51 & QPOC \\
LMC-CEP-0144 & 1O & 6091.7275 & 4.61904(4) & 4.61892(5) & 5.8e-07 [4.4$\sigma$] & 13.70 &  - \\
LMC-CEP-1509 & 1O & 4202.9424 & 4.533075(8) & 4.533143(9) & 1.3e-07 [14$\sigma$] & 13.34 & - \\
LMC-CEP-2669 & F & 4965.1949 & 4.1099263(21) & 4.1099228(31) & -4.1e-09 [1.6$\sigma$] & 14.30 & - \\
LMC-CEP-1369 & F & 5301.1134 & 4.0746625(12) & 4.0746725(20) & 8.2e-09 [6.4$\sigma$] & 14.35 & - \\
LMC-CEP-0491 & F & 5432.4589 & 3.9792895(23) & 3.9792634(30) & -4.1e-08 [13$\sigma$] & 14.47 & - \\
LMC-CEP-0224 & F & 5653.9554 & 3.779639(5) & 3.779641(6) & 3.4e-09 [0.6$\sigma$] & 14.47 & LTTE \\
LMC-CEP-2994 & F & 4945.5601 & 3.703564(4) & 3.703560(4) & -3.8e-08 [6.4$\sigma$] & 14.47 & - \\
LMC-CEP-2208 & F & 4571.4092 & 3.5679932(16) & 3.5679992(22) & 7.2e-09 [3.8$\sigma$] & 14.13 & - \\
LMC-CEP-0160 & F & 5529.5092 & 3.2012952(15) & 3.2013016(19) & 1.1e-08 [5.9$\sigma$] & 14.77 & - \\
LMC-CEP-0889 & F & 4292.0659 & 3.1683767(12) & 3.1683719(14) & -1e-08 [7.0$\sigma$] & 14.83 & LTTE \\
LMC-CEP-0286 & F & 4829.1423 & 3.1335110(18) & 3.1335111(18) & 2.4e-09 [0.9$\sigma$] & 15.04 & LTTE \\
LMC-CEP-3037 & 1O & 5096.1287 & 3.083785(5) & 3.083809(6) & 9.6e-08 [12$\sigma$] & 14.54 & pECL \\
LMC-CEP-1711 & F & 4636.4767 & 2.9189493(15) & 2.9189502(21) & 1e-09 [0.6$\sigma$] & 14.68 & QPOC \\
LMC-CEP-2605 & 1O & 5294.9028 & 2.4194868(32) & 2.419488(5) & 1.2e-09 [0.3$\sigma$] & 14.01 & - \\
LMC-CEP-0110 & 1O & 5010.3116 & 2.1200434(20) & 2.1200490(23) & 1.7e-08 [5.3$\sigma$] & 14.78 & QPOC \\
LMC-CEP-0231 & 1O & 4909.8171 & 1.8993909(24) & 1.8993917(25) & 3.7e-09 [1.0$\sigma$] & 14.57 & - \\
LMC-CEP-2237 & 1O & 4440.6839 & 1.7933278(11) & 1.7933269(11) & 6.2e-09 [3.4$\sigma$] & 14.74 & - \\
LMC-CEP-0334 & 1O & 5631.8010 & 1.7219995(19) & 1.7220008(24) & 2.1e-09 [0.9$\sigma$] & 14.75 & - \\
LMC-CEP-2748 & 1O & 5007.4531 & 1.6514256(13) & 1.6514072(18) & -2.3e-08 [15$\sigma$] & 15.13 & - \\
LMC-CEP-0397 & 1O & 4882.8768 & 1.4772404(6) & 1.4772401(6) & -3.3e-09 [3.9$\sigma$] & 15.23 & - \\
LMC-CEP-2590 & 1O & 4509.3202 & 1.2057281(7) & 1.2057368(7) & 2.9e-08 [36$\sigma$] & 15.72 & - \\
LMC-CEP-2471 & 1O & 5353.4157 & 0.9605283(6) & 0.9605283(7) & 8.8e-12 [0.0$\sigma$] & 16.24 & - \\
LMC-CEP-1347 & 1O/2O & 4695.3526 & 0.69000824(17) & 0.69001053(22) & 3.2e-09 [16$\sigma$] & 16.39 & QPOC \\
LMC-CEP-3287 & 1O/2O & 3342.4313 & 0.6529152(10) & 0.6529149(10) & 1.1e-08 [3.5$\sigma$] & 16.44 & - \\
LMC-CEP-1152 & 1O/2O & 4063.8655 & 0.60684297(13) & 0.60684260(14) & -1.2e-09 [7.8$\sigma$] & 16.36 & - \\
LMC-CEP-2123 & 1O & 5306.5891 & 0.54304670(25) & 0.54304476(32) & -3.8e-09 [11$\sigma$] & 16.65 & LTTE \\
LMC-CEP-2583 & 1O & 5356.1235 & 0.4103540(5) & 0.4103536(6) & -9.5e-10 [1.4$\sigma$] & 17.63 & - \\
LMC-CEP-1662 & 1O & 4063.3731 & 0.31516306(6) & 0.31516315(6) & 2.4e-10 [3.5$\sigma$] & 17.48 & - \\
LMC-CEP-3370 & 1O & 5078.2264 & 0.26924335(11) & 0.26924375(13) & 5.6e-10 [5.1$\sigma$] & 18.01 & - \\
LMC-CEP-2705 & 1O & 4816.0636 & 0.26454063(7) & 0.26454191(10) & 1.6e-09 [19$\sigma$] & 17.21 & - \\
\enddata
\tablecomments{Two ephemerides are presented, assuming either a constant period ($P$) or a linear period change ($dP/dt$). The same reference time ($T_0$, maximum brightness at I-band for a constant $P$) is used for both. Errors in the last digits are given in parentheses. For $dP/dt$ the significance in sigmas is given in brackets. In remarks: LTTE means a presence of a periodic light travel-time effect, QPOC - quasi-periodic cyclic variation in the O-C diagrams, pECL - probably an eclipsing system.}
\end{deluxetable*}

\subsection{O-C analysis} \label{sec:oc}

The supposed orbital motion of our binary Cepheid candidates should result in a slight change of its distance from the observer. Due to the light-travel time effect (LTTE; \citealt{Irwin_1959AJ_LTTE,Borkovits_2015MNRAS_LTTEform}) such a change produces periodic phase shifts in the Cepheid pulsational variability (see examples in \citealt{Udalski_2015_BinCep_LMCSMC,Plachy_2020_TESS_CEPHEIDS}). This effect can be described as:

$$ \Delta_{LTTE} = -\frac{a_{cep}\sin i}{c} \frac{(1-e^2)\sin(\nu+\omega)}{1+e\cos\nu}, $$

\noindent where $a_{cep}$ is the semi-major axis of the Cepheid orbit, $i$ is the orbit inclination, $e$ the eccentricity, $\nu$ is the true anomaly and $\omega$ the argument of periastron, and $c$ is the speed of light. The orbital period $P_{orb}$ and the reference time $T_0$ enters the equation through the true anomaly.

In general, we could expect this kind of phase modulation to be superimposed on a secular evolutionary period variability, which can be easily described with a linear period change. Unfortunately, Cepheids are known to show erratic period variations \citep{Poleski_2008AcA_CEP_PERCHANGE,Suveges_2018AA_CepPerMod}, which are interpreted, for example, as originating from the general instability of the light-curve shape \citep{Derekas_2012MNRAS_OC_VAR} or more specifically, as an effect of convective hot spots \citep{Neilson_2014AA_PER_JITTER}.

Nevertheless, we performed the O-C analysis for all Cepheids in our sample, calculating instantaneous phase shifts along the collected photometric observational data in regard to the ephemeris from Table~ \ref{tab:puls_ephem}, both taking and not taking the linear period change into account. We then looked for any sign of the light-travel time effect due to the supposed binary motion of Cepheids.
As the amplitude of the phase variability in such analysis depends on the size of the projected orbit and the precision depends on the pulsation period, such study works well only for long orbital periods ($\gtrsim$800 days) and Cepheids with short pulsation periods ($\lesssim$3 days). Assuming all of our candidates are binaries with typical periods of 300 days to several years, we expected a clear detection of LTTE only in a few of them, especially taking into account that it can be dominated by the intrinsic (or apparent) period change of Cepheids.

\begin{deluxetable}{cccccc}
%\tablenum{1}
\tablecaption{Orbital parameters from the O-C data\label{tab:oc_orb}}
%\tablewidth{0pt}
\tablehead{
\colhead{OGLE ID} & \colhead{$P_{orb}$} & \colhead{$T_0$} & \colhead{$a_{cep}\sin(i)$} & \colhead{ecc} & \colhead{flags}\\
\colhead{} & \colhead{[days]} & \colhead{[days]} & \colhead{[$R_\odot$]} & \colhead{} & \colhead{}
}
%\decimals
\startdata
LMC-CEP-0224 & 1000 & 4710  & 350 & $\sim$0.35 & M \\
LMC-CEP-0286 &  726 & 4700  & 150 & $\sim$0.1  & M \\
LMC-CEP-0889 & 2533 & 4975  & 950 & 0.0        & - \\
LMC-CEP-2123 & 1720 & 5330  & 480 & $\sim$0.4  & M \\
\enddata
\tablecomments{Parameters from the LTTE fit to the O-C data for Cepheids with periodic phase variability. The full fit was made only for LMC-CEP-0889. For the rest, some parameters were fitted manually (M), because of the strong local O-C deviations.}
\end{deluxetable}

\begin{figure}
    \begin{center}
        \includegraphics[width=1.01\linewidth]{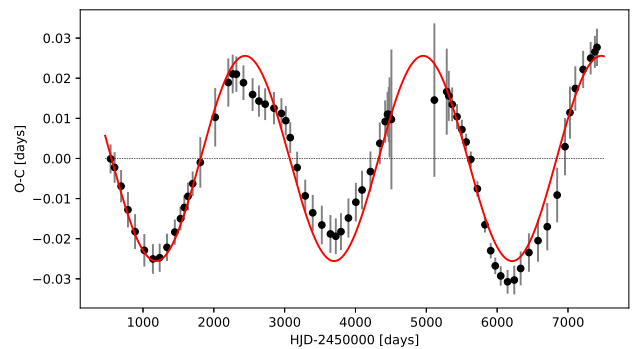}
    \end{center}
\caption{O-C diagram for OGLE-LMC-CEP-0889 ($P=3.17d$) using a non-linear ephemeris from Table~\ref{tab:puls_ephem}. A clear LTTE modulation due to binary motion is present. Red line is the best fit using the orbital parameters from Table~\ref{tab:oc_orb}.}
\label{fig:ltte0889}
\end{figure} 

Indeed, we have one perfectly clear detection of the LTTE due to a binary motion (see Fig.~\ref{fig:ltte0889}) with a period of about 2500 days, and three others for which a clear periodic ($P_{orb}$ from 700 to 1700 days) behavior was observed but with smaller amplitudes. The data for all four systems are given in Table~\ref{tab:oc_orb}.
The orbit of OGLE-LMC-CEP-0889 seems almost circular and quite extended. Its orbital period of about 7 years is probably one of the longest in the sample. The orbits of the other three systems have rather eccentric orbits, but the quality of the fit does not permit to determine it quantitatively. Their orbital periods are about 2, 3 and 5 years. From the data obtained for OGLE-LMC-CEP-0889, we calculated the mass function $f(m) = 1.83 M_\odot$, which suggests a rather high mass of the companion and a high inclination. Conservatively assuming the maximum Cepheid mass of $4.5 M_\odot$ and the inclination of 90 degrees, we obtain the minimum mass ratio of 1.27 for this system. If our assumption of the similar evolutionary stage stands, such a high mass ratio will point to the companion being a merger, similar to OGLE-LMC-CEP-1812, but with the function of the components reversed.

For four objects more, we could detect a cyclic quasi-periodic (tentative periods from 1 to 4 years) behavior, but additional strong erratic changes make these detections more ambiguous. These objects are marked with QPOC in Table~\ref{tab:puls_ephem}.  Moreover, for 25 objects we detected some cyclic variations in the O-C diagrams, which however show strong deviations from periodicity across the photometric data timespan. These can be either stochastic intrinsic period changes or a mixture of such with a weak effect of a binary motion. The O-C data for probably eclipsing OGLE-LMC-CEP-3037 do not yield conclusive results but we found a weak indication of a period of about 3000 days.

From this analysis and from our knowledge on Cepheids in eclipsing binary systems, for objects with no detection of LTTE we expect orbital periods from 1 to 4 years unless very low orbital inclinations made the LTTE signal not detectable for longer period systems.

\section{Spectroscopic confirmation}
\label{sec:spectroscopy}

The first goal of our project was to confirm the hypothesis that all Cepheids selected as described in Section~\ref{sec:object_selection} are members of double-lined binaries by means of spectroscopy.
Such confirmation is not an easy task, however. The expected orbital periods are long and the orbits may be highly eccentric, meaning that for a significant percentage of time the separation in radial velocities is low. Comparing to Cepheids in eclipsing binaries, where inclinations are always between 80-90 degrees, in our case inclinations are scattered randomly between 0 and 90 degrees, on average further decreasing the orbital RV amplitudes. For the same reasons, the change of orbital radial velocities during one observing season may, in general, also be not high enough to unambiguously detect the orbital motion given a very limited number of spectra.
This means that until we know the orbital periods, and are able to calculate when the separations of component radial velocities is the highest, we have to use other technique for confirmation of SB2 status of our candidates.

To check our hypothesis we selected a smaller sub-sample of 18 Cepheids brighter than 15.5 mag in the V-band, for which less observing effort is required. Fourteen of selected targets are fundamental mode Cepheids with RV amplitudes of about 50 km/s. Depending on the period their daily average RV change ranges from 10 to 30 km/s. For the four first-overtone Cepheids that have lower RV amplitudes but also shorter periods, the daily RV change ranges from 10 to 20 km/s. As a typical FWHM (full width at half maximum) of a Cepheid line profile is 20 km/s, thanks to pulsational velocity changes, observations for a few consecutive days should guarantee detection of companions to most of the Cepheids.
Moreover, variable pulsation velocity allows to separate the lines of components even for perfectly face-on orbits, where the orbital RV separation is zero all the time.

We have observed our targets for 4 consecutive nights, obtaining from 2 to 4 high-resolution spectra per object with the HARPS instrument mounted at 3.6-meter telescope at the La Silla observatory in Chile. For most of the candidates one spectrum was also taken earlier using a MIKE spectrograph mounted on the Magellan telescope at the Las Campanas observatory in Chile. We used the reduced HARPS data downloaded from the ESO Archive\footnote{http://archive.eso.org}, while the MIKE data were reduced using Daniel Kelson's pipeline available at the Carnegie Observatories Software Repository\footnote{http://code.obs.carnegiescience.edu}.
For the identification of components in the spectra we used the Broadening Function (BF) technique \citep{Rucinski_1992AJ_BF,Rucinski_1999ASPC_BF_SVD} implemented in the RaveSpan code \citep{t2cep098apj2017}. This technique provides narrower profiles than the cross-correlation function method, which helps in the separation of components and increases the chance of detecting the presence of a companion.

\begin{figure}
    \begin{center}
        \includegraphics[width=0.37\textwidth]{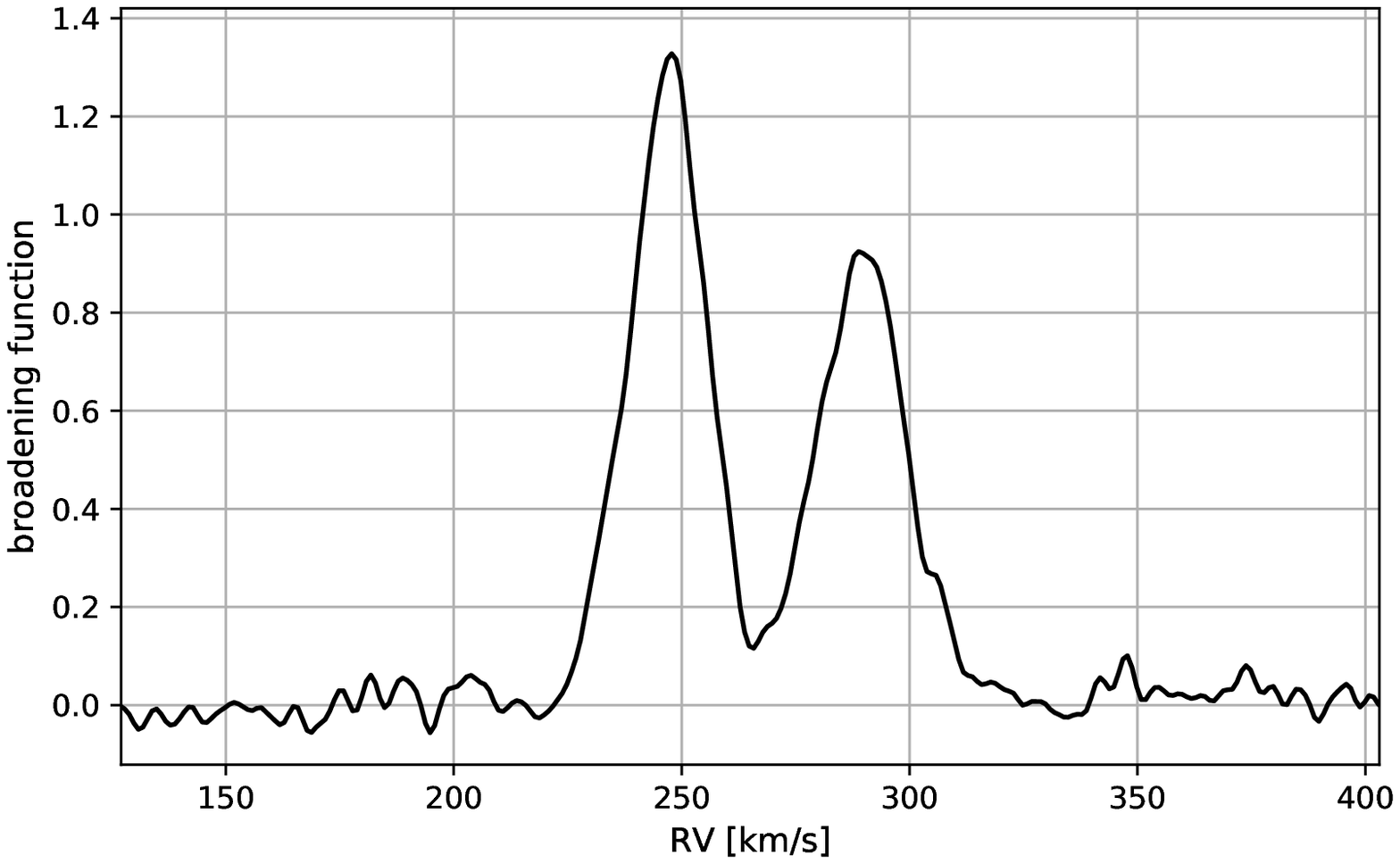}
        \includegraphics[width=0.37\textwidth]{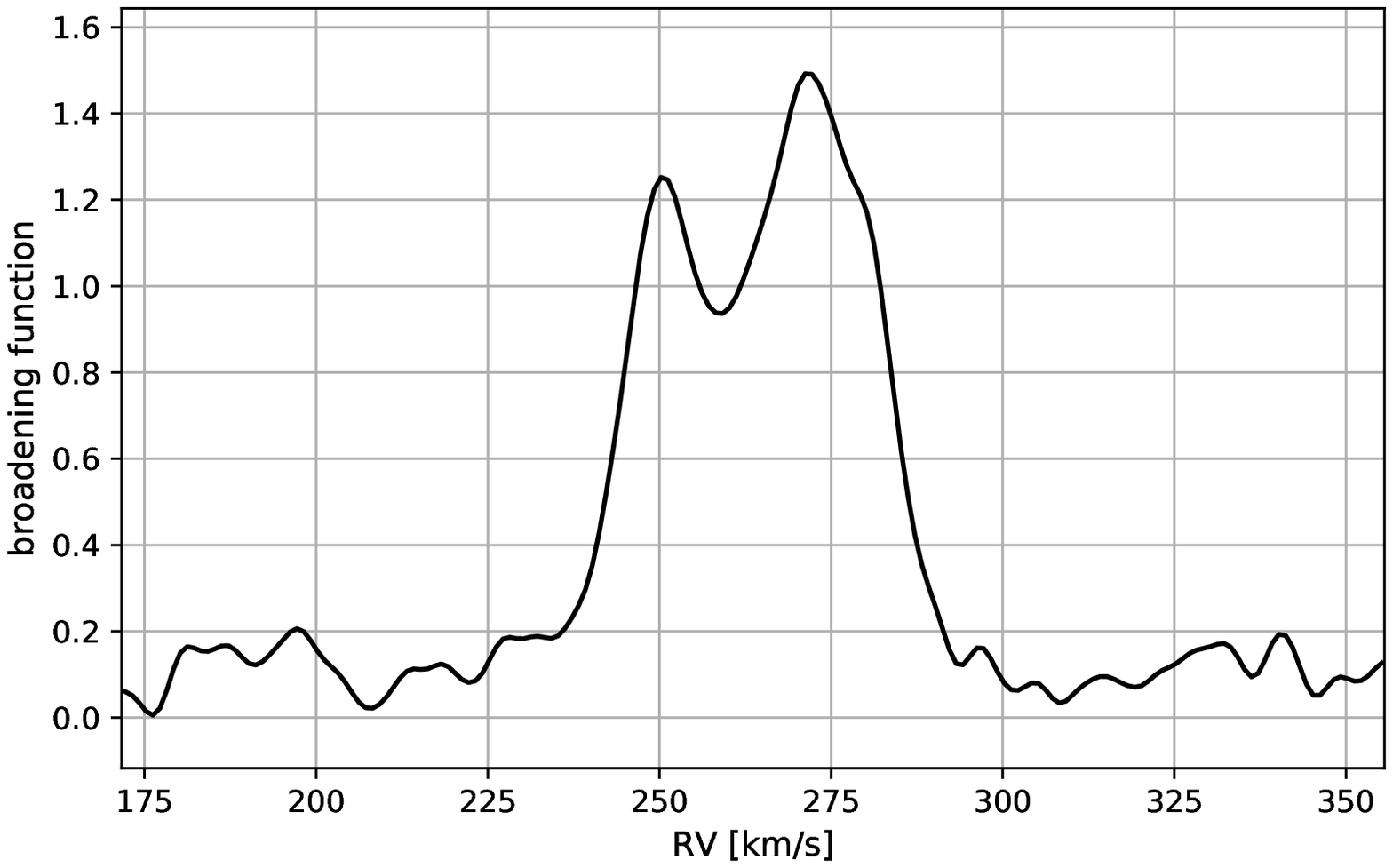}
        \includegraphics[width=0.37\textwidth]{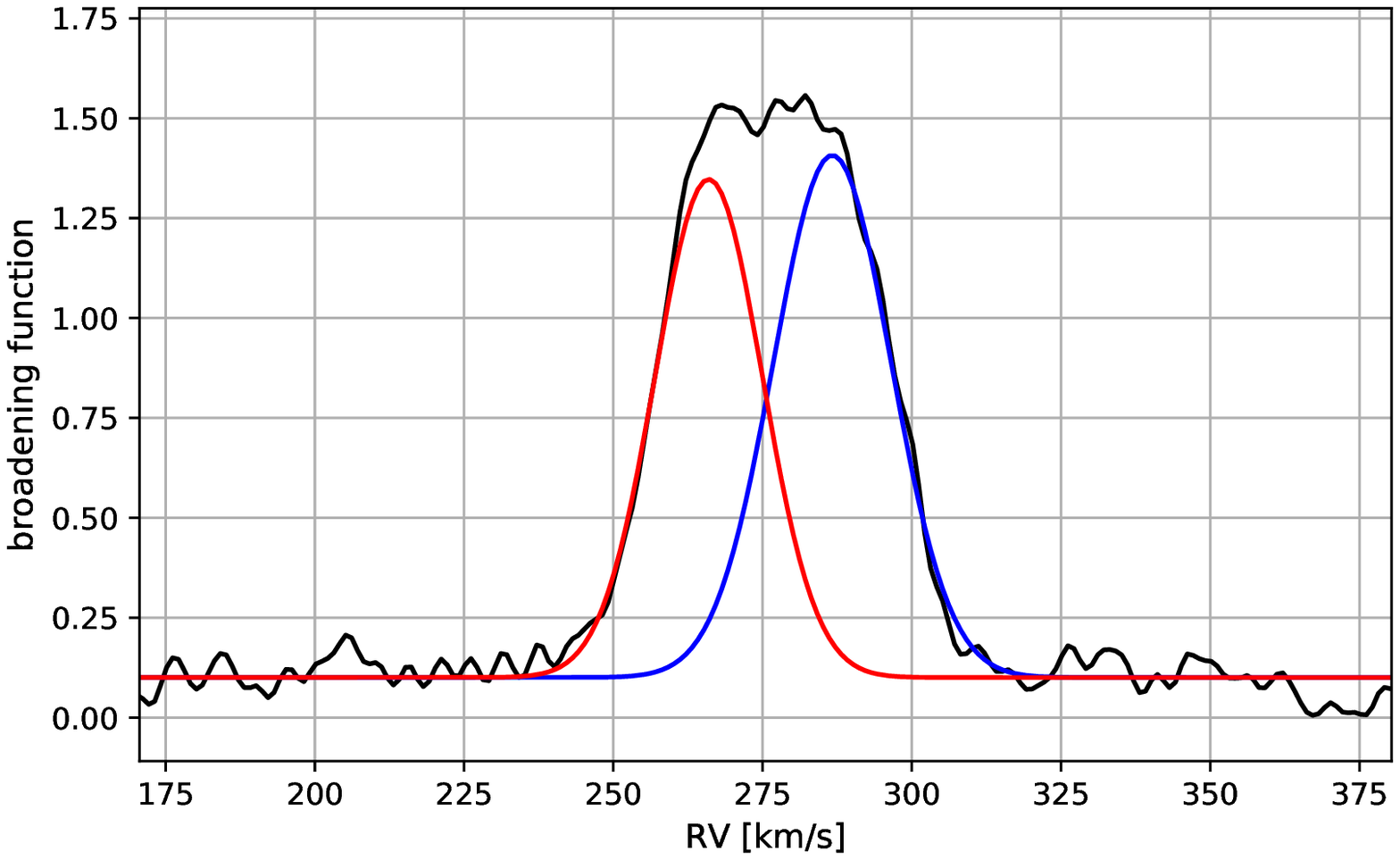}
    \end{center}
\caption{Broadening Function profiles for one example of each class of SB2 status of Cepheids (see Table~\ref{tab:sb2stat}). Left: separated profiles (SEP), middle: 2 peaks visible (2P), right: blended profiles (BP).}
\label{fig:sb2examples}
\end{figure} 

We divided the analyzed objects in four groups, of which three correspond to a positive detection. The negative detection means that two components could not be identified in any of the collected spectra. The objects with a positive detection are those where the profiles of the components are clearly separated (SEP), where the profiles are overlapping, but two peaks are seen (2P) and where the profiles are blended, but their shape can only be explained by a combination of two components (BP). In Fig.~\ref{fig:sb2examples} we show one example for each of these groups.

\begin{deluxetable}{ccccc}
\tablecaption{SB2 status of binary Cepheid candidates \label{tab:sb2stat}}
%\tablewidth{0pt}
\tablehead{
\colhead{OGLE ID} & \colhead{mode} & \colhead{$P_F$ [d]} & \colhead{status} & \colhead{flags} %\\
%\colhead{} & \colhead{} & \colhead{[days]} & \colhead{}
}
\startdata
LMC-CEP-0973 & F & 10.47 & unconf. & ASM \\
LMC-CEP-0751 & F & 9.495 & HProb   & BP \\
LMC-CEP-1472 & 1O & 7.294$^*$ & CLEAR   & 2P \\ 
LMC-CEP-0144 & 1O & 6.903$^*$ & CLEAR   & SEP,cORB\\
LMC-CEP-1957 & F & 6.814 & HProb   & BP \\
LMC-CEP-1509 & 1O & 6.770$^*$ & HProb   & BP \\
LMC-CEP-2096 & F & 5.972 & HProb   & BP \\
LMC-CEP-0837 & F & 5.777 & CLEAR   & SEP,cORB \\
LMC-CEP-0584 & F & 5.725 & CLEAR   & 2P \\
LMC-CEP-2999 & F & 5.406 & CLEAR   & 2P \\
LMC-CEP-1077 & F & 4.654 & CLEAR   & SEP \\ 
LMC-CEP-0835 & F & 4.563 & CLEAR   & 2P \\
LMC-CEP-2669 & F & 4.110 & CLEAR   & 2P \\
LMC-CEP-1369 & F & 4.075 & CLEAR   & 2P \\
LMC-CEP-0491 & F & 3.979 & CLEAR   & SEP \\
LMC-CEP-0224 & F & 3.780 & CLEAR   & 2P \\
LMC-CEP-2208 & F & 3.568 & HProb   & BP \\
LMC-CEP-2605 & 1O & 3.537$^*$ & unconf. & - \\ 
\enddata
\tablecomments{$P_F$ is the pulsation period in case of F-mode Cepheids and fundamentalized period (marked with asterisk) for 1O-mode ones.}
\end{deluxetable}

The final SB2 status of our sub-sample candidates for binary Cepheid candidates is shown in Table~\ref{tab:sb2stat}. Those that belong to the first two positive detection groups (SEP and 2P) are marked as CLEAR cases, those belonging to the third positive group (BP) are marked as High Probability (HProb) cases, and those with a negative detection are marked as unconfirmed. In total we found 16 out of 18 targets to be either clear (11), or high probability (5) cases, and only for two the status is yet unconfirmed. However, for these unconfirmed cases the data were taken unfortunately at very similar Cepheid velocities. The reasons are that OGLE-LMC-CEP-0973 has the longest period in the sample (about 10.5 days), while OGLE-LMC-CEP-2605 has the shortest one (about 2.4 days) and is a low-amplitude 1O-mode Cepheid. For neither of them the extrema of the RV curve were covered. Actually, in some spectra of OGLE-LMC-CEP-0973 we detected profiles asymmetry (ASM) that can be interpreted as coming from a companion, but we prefer to be conservative in this regard.

As explained in Section~\ref{sec:object_selection}, the possibility of an undetected blend is rather excluded for a great majority of these stars.
Moreover the detection of lines of a second component for 90\% of stars confirms that virtually all companions are red, evolved stars. It would be highly improbable that there are practically no Cepheids that are blended with bright early-type main sequence stars.
The above result is thus almost a decisive evidence that even all our candidates may be not only binaries, but very valuable SB2 systems as well. Although, as we noted, we did not expect to detect the orbital motion of Cepheids during one season, for two objects in spectra taken 2 months apart we detected a slight but significant velocity change of the companion, that with high probability can be interpreted as its orbital variation. These two Cepheids have an additional cORB flag in Table~\ref{tab:sb2stat}.

\section{Conclusions and future work}
\label{sec:conclusions}

\begin{figure}
    \begin{center}
        \includegraphics[width=1.01\linewidth]{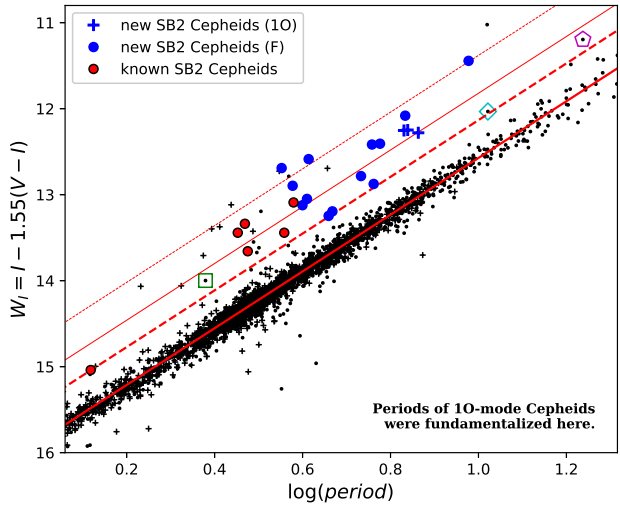} 
    \end{center}
\caption{Similar to Fig.~\ref{fig:perlum} but for periods longer than 1.15 days and showing the newly confirmed Cepheids in SB2 systems. Almost all of them have periods longer than the known SB2 Cepheids, spreading up to about 10 days.}
\label{fig:confirmed}
\end{figure}

In this first paper of the series we presented and proved our hypothesis regarding the double-lined status of overbright Cepheids selected as described in Section~\ref{sec:object_selection}. Up to now we have confirmed this status spectroscopically for 16 out of 18 Cepheids of the sample that are brighter than V=15.5 mag, but the two unconfirmed candidates can still turn to be SB2 binaries once more observations are collected.
Orbital motion was also detected for companions to two Cepheids further strengthening our conclusions about their status. For these two systems rather short orbital periods can be expected. With the 16 confirmed candidates, we have already quadrupled the number of Cepheids in spectroscopic double-lined binaries. In Fig.~\ref{fig:confirmed} one can see that most of these new Cepheids have periods longer than those of known SB2 Cepheids, spreading up to almost 10 days.
The extrapolation of the number of confirmed cases to the whole sample suggests that at least 36 new Cepheids in SB2 systems can be expected once all 41 selected objects are studied. This would mean a huge 7-fold increase in the number of this type of systems.

For four objects (including one already confirmed as SB2) we also found a periodic variability in the O-C diagrams. From these diagrams the first-guess orbital periods of the Cepheids can be derived. Although future observations will be necessary to confirm it, the initial periods will help considerably in establishing the final ones with lower number of spectra collected. The same is true for the Cepheid for which one eclipse was observed. For that system we can also expect the inclination close to 90 degrees, which will further simplify the analysis.

If periods of all four objects showing significant LTTE (due to binary motion) are confirmed, we can learn from them to what extent the stochastic variation of the Cepheid period may affect the LTTE. This may help in the identification of less clear cases and to confirm those where cyclic yet not 100\% periodic behavior was detected (marked QPOC). For one system we could estimate the minimum mass ratio, which points to a merger scenario for the companion. This will still have to be confirmed spectroscopically but it suggests that more such interesting cases can be expected in the whole sample.

The full project dedicated to the LMC sample will take several years to be completed. We decided that it is important to publish these preliminary results as they already have important implications for the interpretation of period-luminosity relations and for our general knowledge of Cepheids and their evolution. For example, we have now a strong clue that overbright Cepheids, that were often being rejected as P-L relation outliers, are just Cepheids with red, luminous (giant) companions.
These conclusions can be likely extended to other pulsating stars with well-defined P-L relations like, for example, Type-II Cepheids or RR Lyrae stars.

\subsection{Future work}

The next step will be to check the double-lined status for the whole sample of 41 Cepheids and to confirm the orbital motion for those that are already identified as SB2. Although already there is very little chance that the candidates showing lines of two components are two unrelated stars, the detection of the Cepheid and companion orbital motion will be the ultimate proof and a firm base for the future mass determination.

In the meantime, once enough data is collected, we will start characterizing the components. We will measure the spectroscopic light ratios and analyze their spectra that will provide first estimates of their temperatures, metallicities and surface gravities. 

We then plan to monitor these systems with the aim to measure the exact orbital period and obtain the full orbital solution, including the measurement of the mass ratio for them. We expect to be able to do so for the great majority of Cepheids within 5 years from now, while several systems will possibly need a few years more to be analyzed. Note that to have good mass ratios, measurements around quadratures are needed, which means that on average 3/4 of the orbital cycle have to be covered (assuming starting observations at random orbital phase). Knowing the mass ratios will already tell us a lot about the evolution, original multiplicity and occurrence of mergers in multiple systems of intermediate-mass stars.
Moreover, for non-eclipsing SB2 systems lower mass limits ($M \sin^3 i$) can also be derived directly. While for a single system such knowledge has a limited value, a statistically significant sample brings a multitude of possibilities.

We can also use prior knowledge to put much stronger constraints on masses. The best way would be to use the distance to the system (from the accurate distance to the LMC; \citealt{Pietrzyn_2019_LMC_1perc}) and the expected brightness of the Cepheid (from the P-L relation) or the spectroscopic light ratio, to calculate the luminosity of both components. Using temperatures obtained from colors or spectra we can then determine the radii and use the period-mass-radius relation or directly the pulsation theory to determine the mass of the Cepheid \citep{t2cep098apj2017,allcep_pilecki_2018}. Having the mass ratio, a mass of the companion can also be calculated, providing the full characterization of the system.

In the long run our study will bring firm mass estimates for a large sample of Cepheids (about 4 times more numerous than available today), including long-period, high-mass Cepheids for which lack of data is the most severe, and probing the uncertain low-mass end at the same time. A similar stage of evolution of both components implies mass ratios close to one. Any significant deviation from unity will be extremely interesting, indicating a very probable merger event in the Cepheid past. And this is the only way merger Cepheids can be unambiguously detected (apart from dubious chemical composition peculiarities) and characterized.

We have also already identified a significant number of similar Cepheid candidates in the Small Magellanic Cloud. Moreover, Gaia eDR3 provides sufficient precision to look for such candidates also in the Milky Way. Extension of our project to these two galaxies will further increase the number of Cepheids in SB2 systems leading towards an order of magnitude improvement in our knowledge of Cepheids physical properties.

\acknowledgments
{\small The research leading to these results has received funding from the European Research Council (ERC) under the European Union's Horizon 2020 research and innovation program (grant agreement No 695099) and from the Polish National Science Center grant MAESTRO UMO-2017/26/A/ST9/00446. We also acknowledge the grant MNiSW DIR/WK/2018/09.
W.G. and G.P. gratefully acknowledge support from the BASAL Centro de Astrof{\'i}sica y Tecnolog{\'i}as Afines (CATA) AFB-170002. W.G. also acknowledges support from the Millenium Institute of Astrophysics (MAS) of the Iniciativa Cientifica Milenio del Ministerio de Economia, Fomento y Turismo de Chile, project IC120009. M.T. acknowledges financial support from the Polish National Science Centre grant PRELUDIUM 2016/21/N/ST9/03310.}

{\small This work is based on observations collected at the European Southern Observatory under ESO programme 106.21GB.003. We also thank Carnegie, and the CNTAC for the allocation of observing time for this project. We would like to thank the support staff at the ESO La Silla observatory and at the Las Campanas Observatory for their help in the remote observations.}
{\small This research has made use of NASA's Astrophysics Data System Service.}

\software{
\texttt{RaveSpan} \citep[][\url{https://users.camk.edu.pl/pilecki/ravespan/}]{t2cep098apj2017}
}



\bibliography{ccbincand1}{}
\bibliographystyle{aasjournal}



\end{document}